\documentclass[linenumbers]{aastex631}

\usepackage{float}
\usepackage{amsmath}
\usepackage{booktabs} 
\let\tablenum\relax
\usepackage{siunitx}

\begin{document}

\title{Temporal Variations in Asteroseismic Frequencies of KIC 6106415: Insights into the Solar-Stellar Activity from GOLF and Kepler Observations}

\author[0009-0002-0223-6482]{Christopher J. Lombardi}
\affiliation{Center for Computational Heliophysics, New Jersey Institute of Technology, University Heights, Newark, NJ 07102-1982, USA}

\author[0000-0003-0364-4883
]{Alexander G. Kosovichev}
\affiliation{Center for Computational Heliophysics, New Jersey Institute of Technology, University Heights, Newark, NJ 07102-1982, USA}
\affiliation{NASA Ames Research Center, Moffett Field, Mountain View, CA 94040, USA}

\author[0000-0003-2002-0247]{Keitarou  Matsumoto}
\affiliation{Center for Solar-Terrestrial Research, New Jersey Institute of Technology, University Heights, Newark, NJ 07102-1982, USA}

\begin{abstract}
The Global Oscillations at Low Frequencies (GOLF) instrument aboard the Solar and Heliospheric Observatory (SOHO) mission has provided over two decades of continuous, high-precision data, enabling detailed measurements of the Sun's oscillation frequencies. These oscillations, analyzed through Doppler velocity shifts, offer invaluable insights into the Sun’s internal structure and dynamics by applying methods of helioseismology. This methodology has been extended to the study of stars beyond the Sun, using data from various space missions. In particular, NASA's Kepler mission, in operation from 2009 until 2018, observed over 500,000 stars, analyzing variations in brightness over time and creating a vast database of photometric data to study. This investigation focuses on the solar-type star KIC 6106415, comparing its oscillation frequencies with those derived from the GOLF data. By analyzing frequency patterns and mode lifetimes, we explore similarities and differences in internal structures, stellar evolution, and magnetic activity cycles between KIC 6106415 and the Sun. Our analysis reveals that KIC 6106415 exhibits starspot numbers similar to the Sun, peaking at an estimated 175, consistent with its faster rotation rate. The data suggest that KIC 6106415 may have shorter magnetic activity cycles than the Sun, reinforcing the link between stellar rotation and magnetic field generation in solar-type stars.

\end{abstract}

\section{Introduction} \label{sec:intro}{Although significant progress has been made in observing the external layers of stars, their interior properties remained largely mysterious until the advent of helioseismology in the 1990s \citep[e.g.][]{Kosovichev2011}. By studying sound waves trapped within the Sun, this revolutionary technique enables the quantification of key stellar parameters, including mass, age, chemical composition, and radius. When applied to stars beyond the Sun, this approach is known as asteroseismology. It serves as an invaluable tool for placing the Sun in a broader stellar context, enhancing our understanding of stellar evolution and dynamics, while also providing insights into the behavior of solar-type stars.

NASA's Kepler mission, originally designed to detect exoplanets, has provided extensive datasets of light curves, representing the flux of stellar brightness as a function of time \citep{Borucki2010}. Many of the observed stars are classified as solar-type, with similar mass, radius, and chemical composition to the Sun. Investigating these stars is critical for placing the Sun within a broader astrophysical framework. Preceding the Kepler mission, the Global Oscillations at Low Frequency (GOLF) experiment \citep{GOLF} aboard the Solar and Heliospheric Observatory (SOHO) was commissioned to study both gravity and acoustic modes on the solar disk using a resonant scattering photometer \citep{Gabriel1995}. Acoustic modes, in particular, play a central role in probing the internal structure of stars, as they are more sensitive to the outer convective regions and are more easily detectable due to their higher frequency values \citep{Lopes2001}.

In the Sun and other similar stars, acoustic modes are stochastically excited by convective motions in the outer regions. Temporal variations in mode frequencies can be observed for different radial orders, $\mathit{n}$, and angular degrees, $\ell$, due to their sensitivity to different stellar layers. Notably, higher-frequency modes in the Sun exhibit larger shifts compared to the smaller shifts seen in lower-frequency oscillations \citep[e.g.][]{Kosovichev2011}.

In this study, we present an analysis of the solar-type star KIC 6106145, utilizing three years of high-precision photometric observations from the Kepler mission \citep{MAST2023}. The light curves of KIC 6106145 reveal evidence of magnetic activity in the form of stellar flares, suggesting a dynamic outer layer resembling the Sun's. These flares may play a role in the stochastic excitation of oscillation modes, potentially influencing the observed temporal variations in mode frequencies. Our results highlight that higher-frequency oscillation modes in KIC 6106145 exhibit more pronounced temporal variations compared to lower-frequency modes, mirroring behavior observed in the Sun. This finding underscores the connection between surface magnetic activity and stellar oscillations, offering valuable information on the physical processes that occur in the outer layers of solar-type stars \citep{Broomhall2015}.

\section{Kepler Light Curves} \label{sec:Light Curves}

The Kepler mission, renowned for its high-precision photometric observations, provided a wealth of data essential for studying stellar oscillations and magnetic activity. For this study, the Mikulski Archive for Space Telescopes (MAST) database was queried to identify stars with parameters similar to the Sun, such as mass, radius, effective temperature, and metallicity. From this search, three years of pre-conditioned simple aperture photometry (PDCSAP) light curves were downloaded for detailed analysis of KIC6106415, one of which is shown in Figure 1 \citep{MAST2023}. These light curves differ from the standard SAP flux in that they aim to eliminate planetary signals and other systematic noise in the form of cotrending basis vectors \citep{Reinhold2013}. The data used for this study were obtained at a 1-minute cadence, allowing the identification of oscillation periods comparable to those of the Sun, which are on the order of 5 min. The Python software package, AltaiPony, allowed for identification of flare activity in the time series. Data points in each observation window were compared with a known flare model, with points above a chosen threshold highlighted to indicate possible areas of interest (see Figure 2) \citep{Ilin2021}.

To prepare the data for frequency analysis, the light curves were detrended to remove long-term variations caused by instrumental effects or stellar variability unrelated to oscillations. A Savitzky-Golay filter, also implemented through the AltaiPony library, was applied to preserve oscillatory signals while eliminating spurious noise \citep{Ilin2021}. The data were also corrected for outliers and normalized to easily visualize the deviations. To enhance frequency resolution, the light curves were stitched together into 90-day observation periods, comprising each Kepler quarter, with a 45-day overlap between consecutive quarters \citep{MAST2023} (see Figure 3). This approach improved the ability to resolve closely spaced frequencies in the subsequent spectral analysis.

\begin{figure*}[h] 
    \centering
    \includegraphics[width=\textwidth]{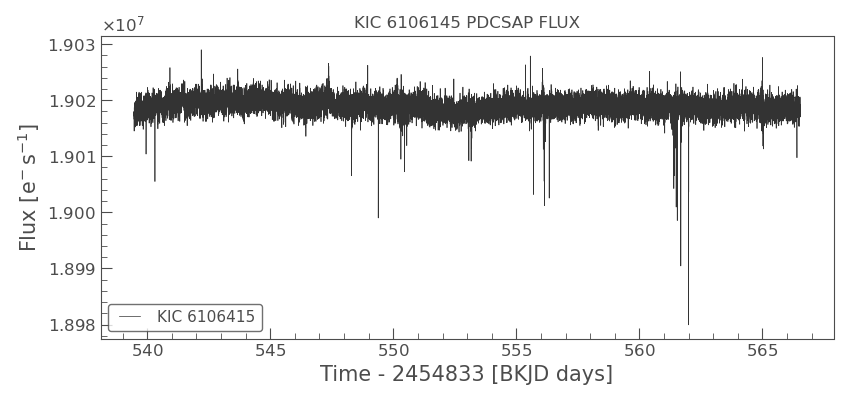} 
    \caption{Example of the light curve, before corrections applied, downloaded from MAST database for 30-day observation period.}
    \label{fig:example}
\end{figure*}

\begin{figure*}[h] 
    \includegraphics[width=\textwidth]{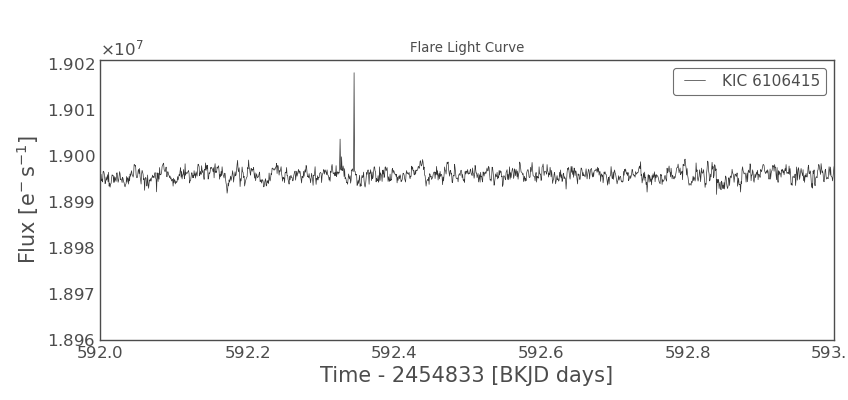} 
    \caption{The light curve highlighting the stellar flare activity.}
    \label{light curve}
\end{figure*}

\begin{figure*}[h] 
    \centering
    \includegraphics[width=\textwidth]{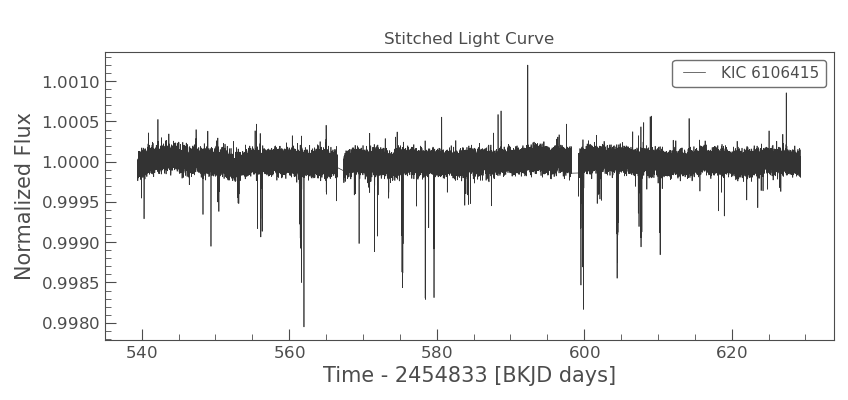} 
    \caption{Light curve stitched together in a 90 day observation window. The data have been normalized to better show flux variations away from the mean value.}
    \label{stitched light curve}
\end{figure*}

The processed light curves were then transformed into the frequency domain using the Lomb-Scargle periodogram provided in the Python library, Lightkurve \citep{2018ascl.soft12013L}. This method was chosen because of the uneven sampling of the data, and to avoid using interpolation methods to make the time series more consistent with the traditional Fourier transform \citep{VanderPlas2018}. Interpolation is known to introduce aliasing in the frequency peaks of the resulting periodogram. An oversample factor of 10 was used in calculating these periodograms to more accurately resolve closely spaced peaks in the spectrum \citep{VanderPlas2018}. This value was chosen based on the frequency resolution, calculated from $\mathit{1/T}$, where $\mathit{T}$ represents the total observation time. By dividing this value by the oversampling factor we attain the grid spacing in the frequencies, which was $\mathit{0.013} \mu$Hz for this data \citep{VanderPlas2018}. Maximum and minimum values were set to highlight the p-mode section of the frequency domain. These frequency spectra were critical for identifying the oscillation modes of KIC 6106145 and provided insights into the star's magnetic activity and oscillatory behavior. By leveraging this high-cadence data, we were able to analyze the stellar oscillations of KIC 6106145, furthering our understanding of solar-type stars.\\

\begin{figure*}[h] 
    \centering
    \includegraphics[width=\textwidth]{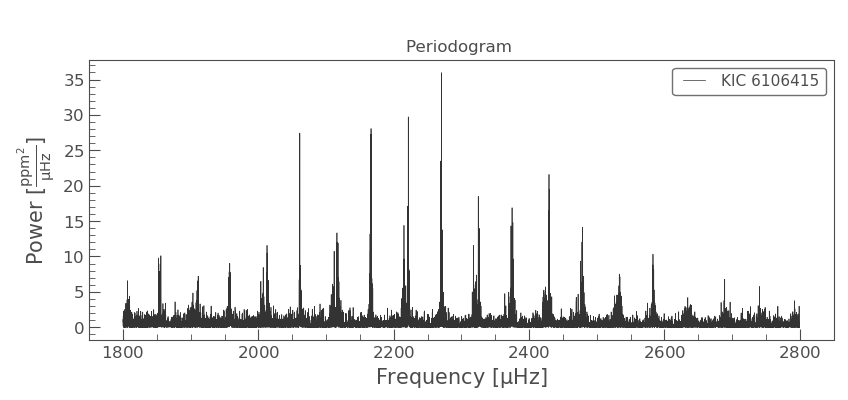}  with your image file
    \caption{Lomb-Scargle periodogram calculated from the above 90-day observation period. Power has been normalized to relative units.}
    \label{Periodogram}
\end{figure*}

\section{Analysis of GOLF Data} \label{sec:GOLF}

The global oscillation frequencies of the p-modes from the solar observations have been recorded for Solar Cycles 22 and 23 \citep{Salabert2011}. The temporal shifts of these frequencies exhibit a strong dependence on magnitude, with higher frequency modes showing larger variations compared to lower frequencies. This is attributed to the differing sensitivities of p-modes to various depths below the solar surface, with higher frequency modes probing shallower layers and lower frequency modes penetrating deeper \citep{Salabert2015}. There is also a strong dependence on $\ell$ value, with the low degree modes probing deeper into interior layers of the star \citep{Demarque1999}. These shifts are indicative of structural changes in the sub-surface layers, such as variations in temperature or the size of the acoustic cavity, and are closely linked to the spatial distribution of the Sun's magnetic field. It has been known for some time that the frequencies also depend heavily on the Sun's 11-year magnetic cycle, with values at a maximum when the Solar activity also reaches it's highest point \citep{Broomhall2015}. \cite{Salabert2011} concluded their analysis midway through Solar Cycle 24, highlighting the weaker activity levels of the cycle and the corresponding smaller frequency shifts compared to Cycle 23. Our subsequent work extended the analysis to include the complete Cycles 23 and 24.

SOHO data was utilized from 1996 until 2022 and, using a similar methodology to \cite{Salabert2011}, we identified that higher angular degree modes ($\ell = 2$) exhibited greater temporal variations compared to lower-degree modes. Figures 5, 6 and 7 show the frequency variations, averaged over all observed $\mathit{n}$ values, for $\ell = 1, 2, 3$ respectively. $\ell=0$ modes were averaged over $\mathit{n = 14-24}$, $\ell=1$ modes were average over $\mathit{n=13-24}$, and $\ell=2$ modes were average over $\mathit{n=14-23}$. The range of frequencies for the $\ell = 2$ modes indicated values between -0.0002 mHz and 0.0002 mHz, corresponding to a 0.0004 mHz fluctuation. This range represents a greater overall variation when compared to the 0.0002 mHz fluctuation for $\ell = 0$ and the approximately 0.00025 mHz change for $\ell = 1$. Again, since higher degree $\ell$ modes are more sensitive to the outer layers of the sun, this finding is consistent with the expectation that they will exhibit greater temporal variations than the lower-degree mode frequencies. 

\begin{figure}[ht!]
\plotone{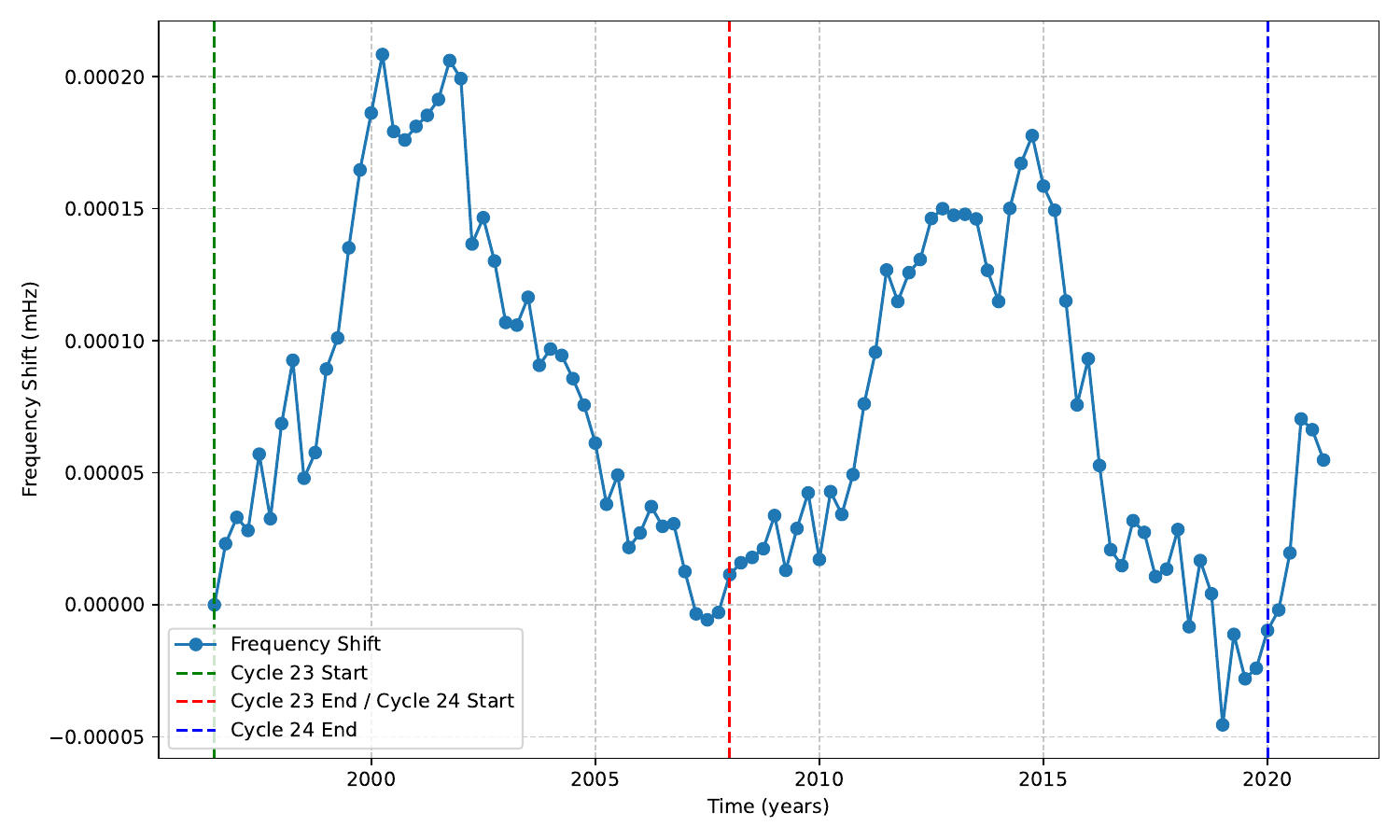}
\caption{$\ell=0$ frequency shifts averaged over all $\mathit{n}$ values. 
\label{L0}}
\end{figure}

\begin{figure}[ht!]
\plotone{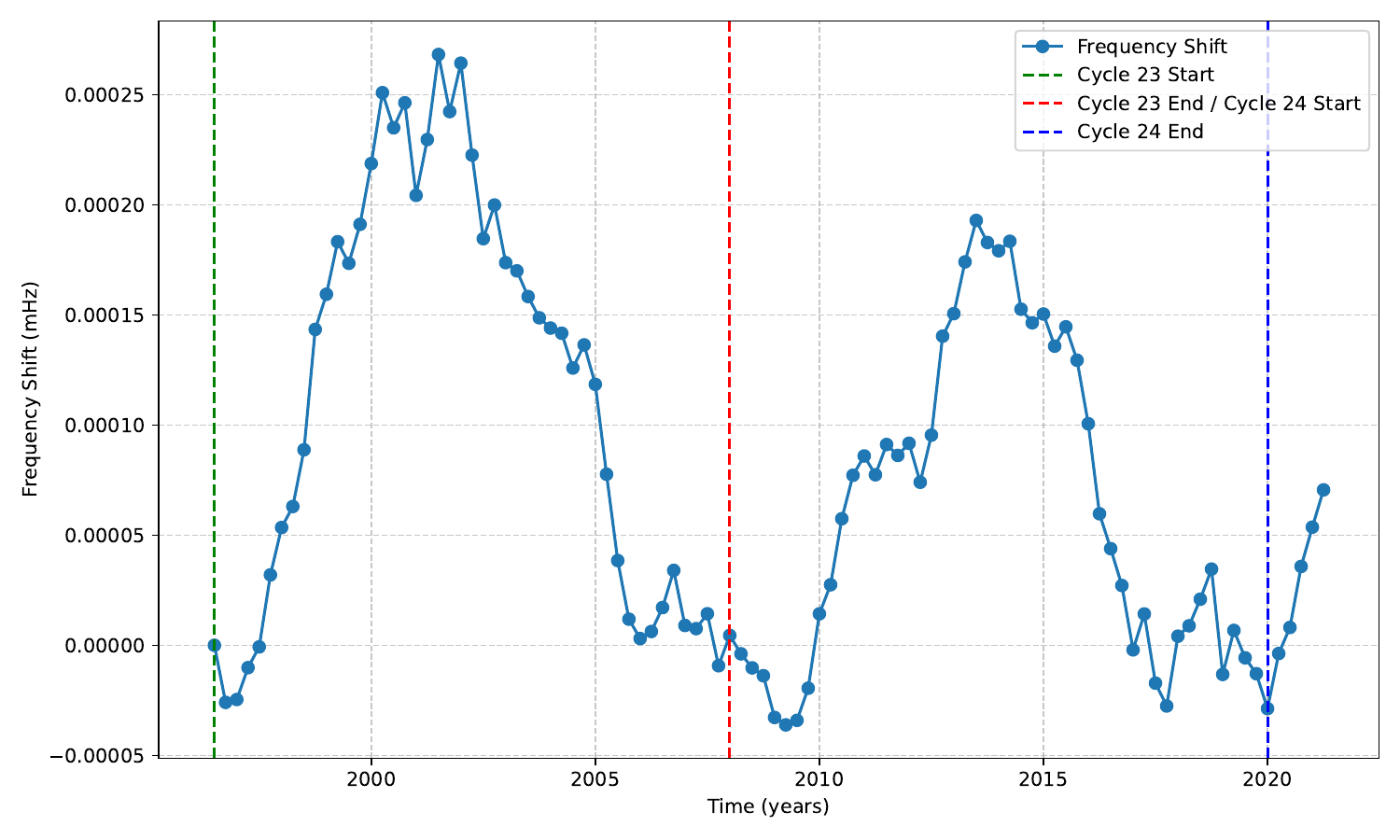}
\caption{$\ell=1$ frequency shifts averaged over all $\mathit{n}$ values. 
\label{L1}}
\end{figure}

This result underscores the potential for more detailed studies of mode dependence variations to further elucidate the effects of magnetic and structural changes in the Sun’s sub-surface layers. By including the full duration of Cycles 23 and 24, our work captures the contrasting activity levels between cycles. The consistent methodological framework applied here enables strong comparisons across cycles, ensuring that observed variations are intrinsic to the Sun and not artifacts of data reduction or analysis techniques. In addition to expanding our understanding of solar oscillations, these results have broader implications for stellar astrophysics. The sensitivity of p-mode frequencies to surface magnetic fields provides a powerful diagnostic tool for studying magnetic activity in solar analogs. This cross-disciplinary approach promises to deepen our understanding of the underlying physics driving stellar variability and evolution.

\begin{figure}[ht!]
\plotone{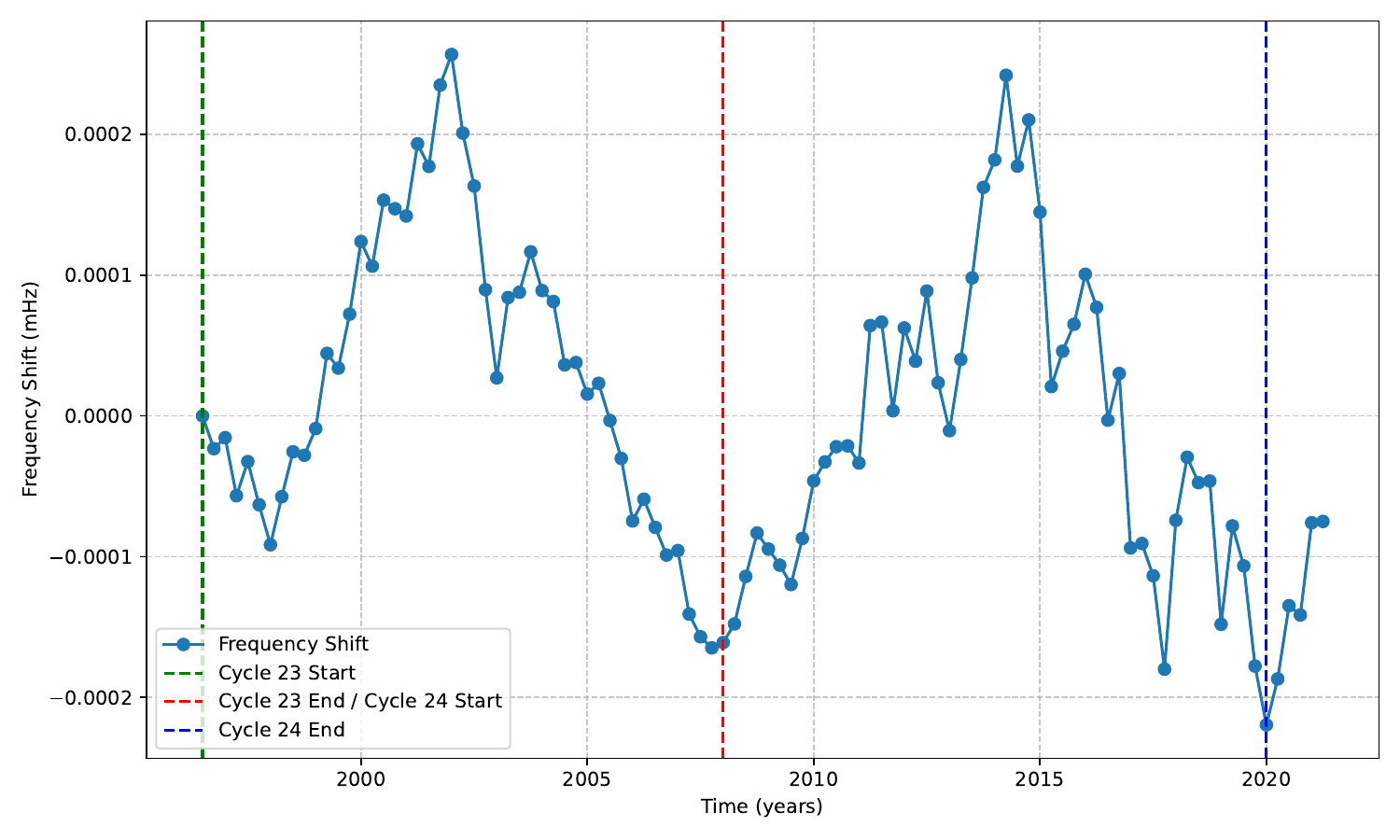}
\caption{$\ell=2$ frequency shifts averaged over all $\mathit{n}$ values. 
\label{L2}}
\end{figure}
\hspace{20mm}
\section{Mode Fitting for KIC 6106415\label{sec:modes}}

The oscillation modes of KIC 6106415 were analyzed using the Apollinaire package, a Python-based tool specifically designed for Bayesian peakbagging in asteroseismology \citep{Breton2022}. Apollinaire employs a Markov Chain Monte Carlo (MCMC) technique to accurately fit the oscillation modes by exploring the parameter space and determining the most probable values for frequency, amplitude, and width. This approach provides a robust statistical framework for quantifying uncertainties in the fitted mode parameters, ensuring reliable results (for more info see \cite{Breton2022}). The Bayesian framework further enables the incorporation of prior knowledge about mode properties, enhancing the precision of the fits and reducing the risk of overfitting.

For this analysis, frequency tables were generated for different radial orders, $\mathit{n}$ and angular degrees, $\ell$, over each observational period. These frequency tables allowed for detailed tracking of temporal variations in the oscillation modes, revealing the impact of magnetic activity and structural changes on the stellar interior. Individual mode frequencies were plotted to visualize their behavior over time, highlighting both short-term fluctuations driven by transient activity and long-term trends associated with stellar magnetic cycles. This approach provided a comprehensive view of how the oscillation modes evolved during the observation period.

Figures 8, 9, and 10 display examples of these frequency plots for chosen $\mathit{n}$ values, with error bars to showcase uncertainties in the calculations. To capture a broader overview, the mode frequencies were averaged across all $\mathit{n}$ and 
$\ell$ values for each observation, and these averages were plotted to assess the overall temporal evolution of the star’s oscillatory properties. This averaging process provided insights into global changes in the star's acoustic cavity and its magnetic environment, particularly during periods of heightened stellar activity. To investigate potential frequency-dependent trends, the frequencies were divided into two regimes: high frequency and low frequency. This division was based on a threshold, approximately halfway between the minimum and maximum observed values, ensuring a balanced analysis across the frequency spectrum. By separating the data into these two regimes, distinct trends in frequency shifts were analyzed, offering a detailed perspective on how the different layers of KIC 6106415's interior respond to magnetic activity, temperature variations, and other stellar dynamics.

\begin{figure}[hbt!]
\plotone{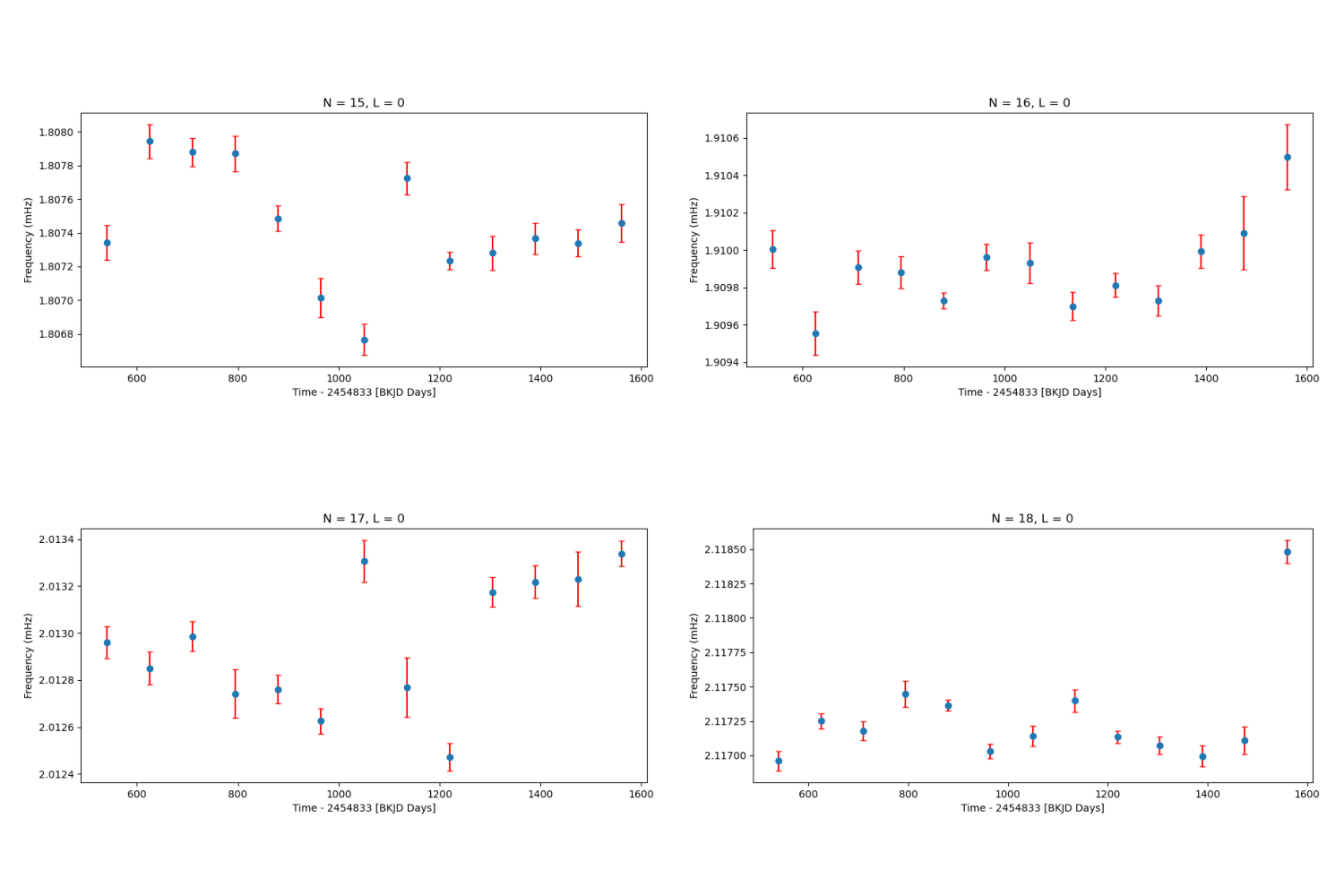}
\caption{Temporal Variations for $\ell = 0$ modes for $\mathit{n= 16, 17, 18, 19}$. 
\label{L0 Golf}}
\end{figure}

\begin{figure}[hbt!]
\plotone{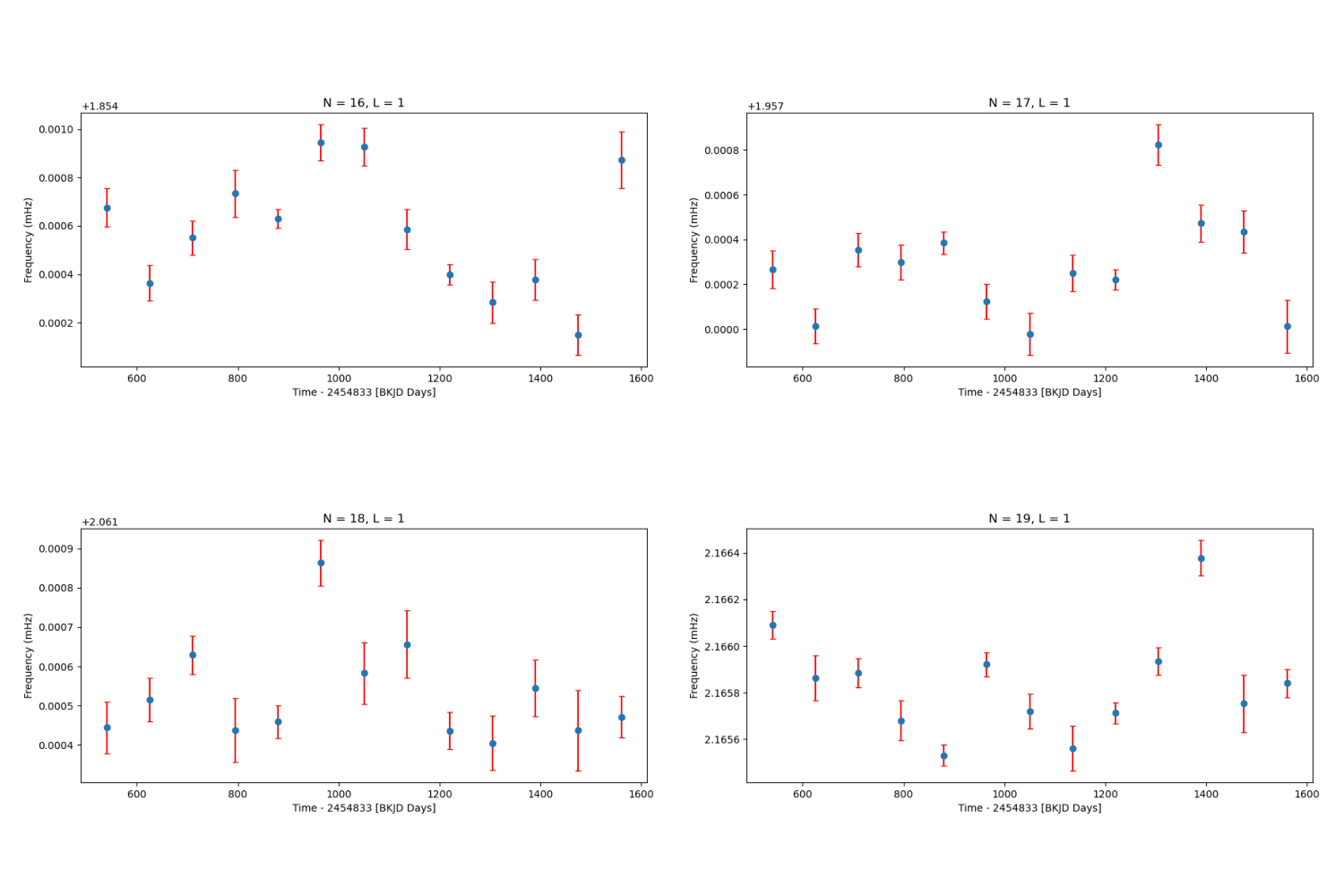}
\caption{Temporal Variations for $\ell = 1$ modes for $\mathit{n = 15, 16, 17, 18}$. 
\label{L1 Golf}}
\end{figure}

\begin{figure}[hbt!]
\plotone{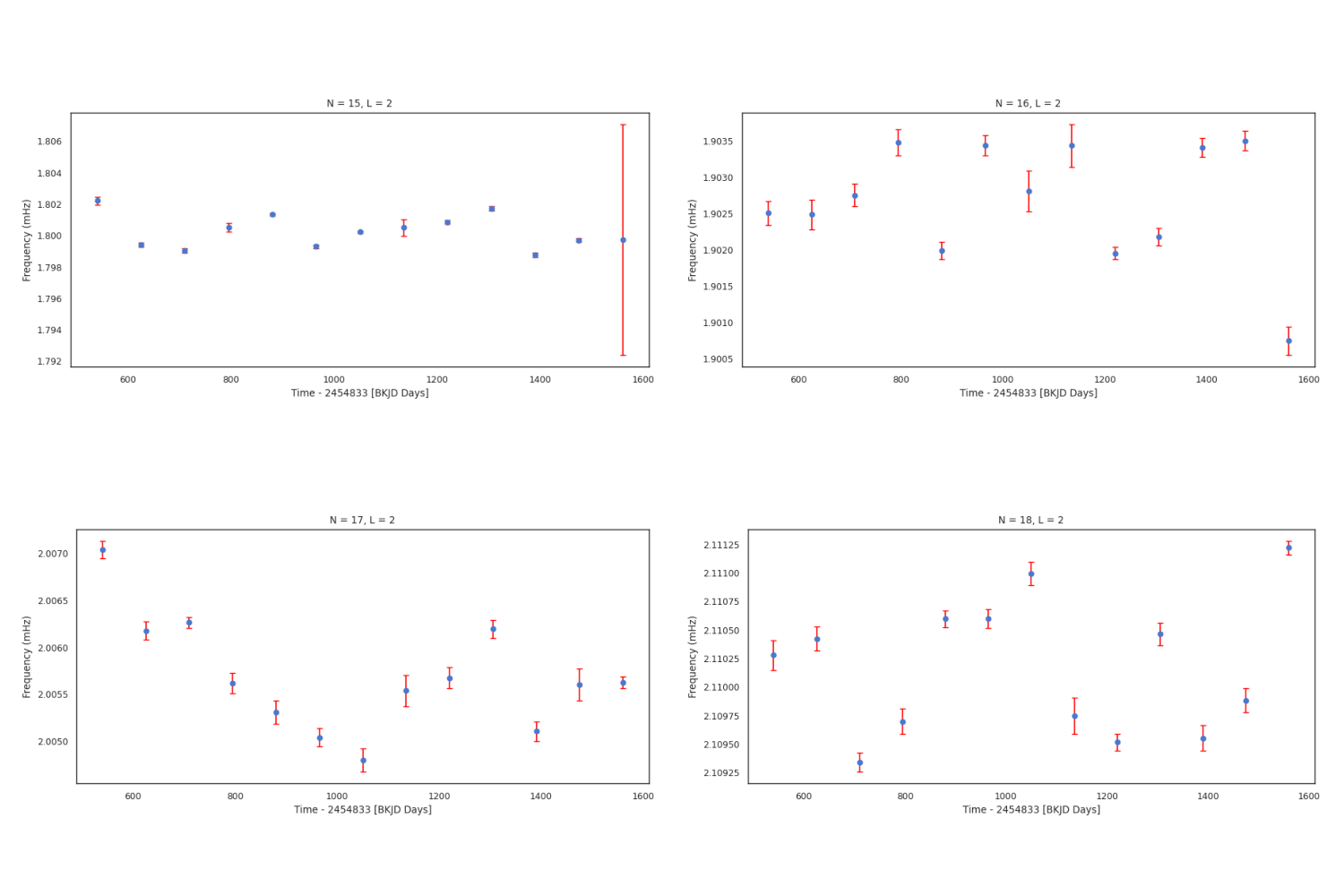}
\caption{Temporal Variations for $\ell = 2$ modes for $\mathit{n = 15, 16, 17, 18
}$. 
\label{L2 Golf}}
\end{figure}

\section{Comparision of Temporal Variations of Oscillation Frequencies\label{sec:Summary}}
The Sun’s oscillation frequencies, as measured by GOLF, exhibit well-documented temporal variations closely tied to solar magnetic activity \citep{GOLF}. During periods of high solar activity, such as solar maximum, higher-frequency modes display larger frequency shifts due to their sensitivity to near-surface magnetic field perturbations and temperature variations. Lower-frequency modes, which probe deeper into the solar interior, exhibit relatively smaller shifts \citep{Salabert2011}. This frequency-dependent behavior serves as a diagnostic for understanding the Sun’s internal dynamics and magnetic activity.

\begin{figure}[h!]
\plotone{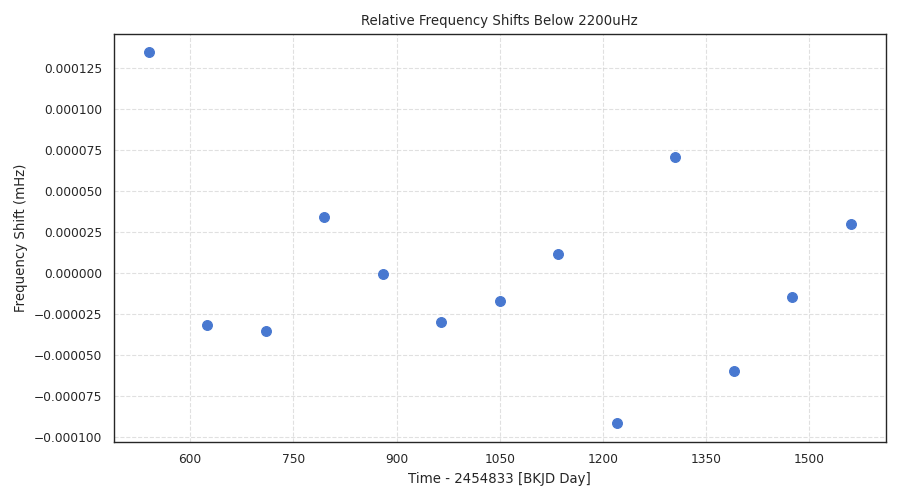}
\caption{Frequency Variations below 2200 $\mu$Hz. Temporal changes in frequency are lower than the corresponding higher frequencies. 
\label{total frequencies below}}
\end{figure}

\begin{figure}[h!]
\plotone{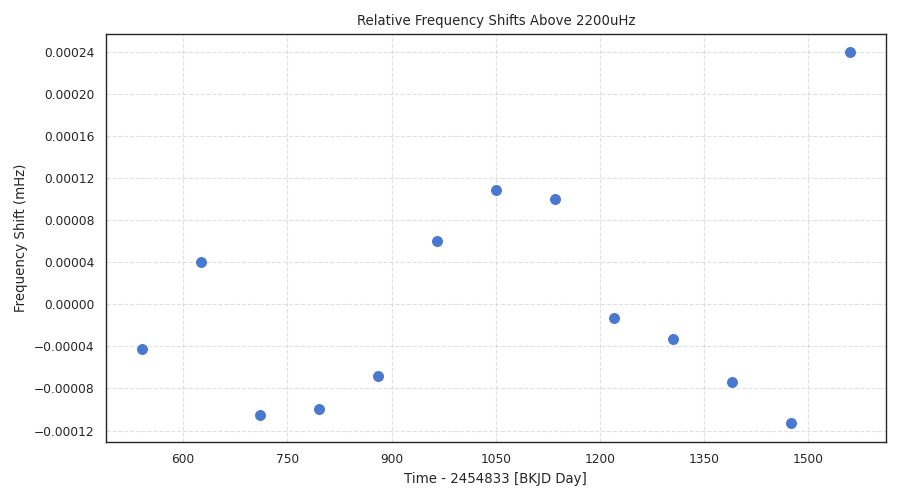}
\caption{Frequency Variations above 2200 $\mu$Hz. 
\label{fig:general}}
\end{figure}

\begin{figure}[b!]
\plotone{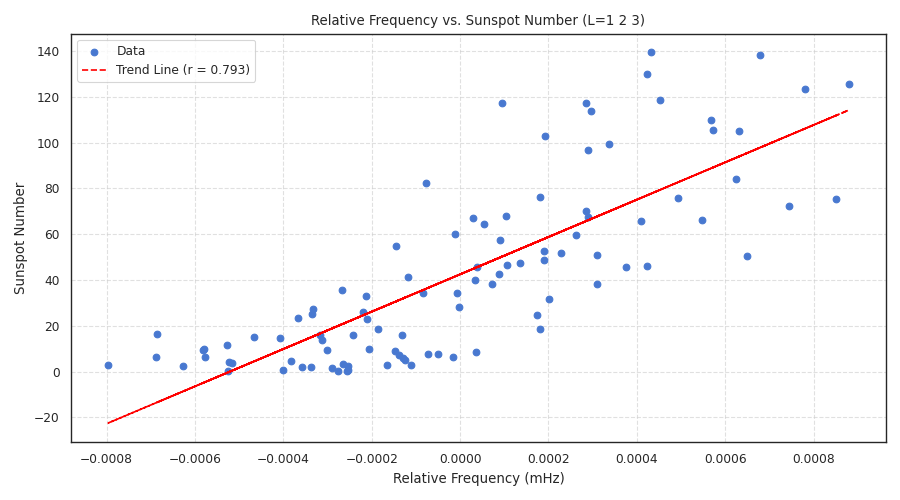}
\caption{Relative Frequency Variations plotted against Sunspot Numbers from GOLF data. 
\label{total frequencies above}}
\end{figure}

\begin{figure}[b!]
\plotone{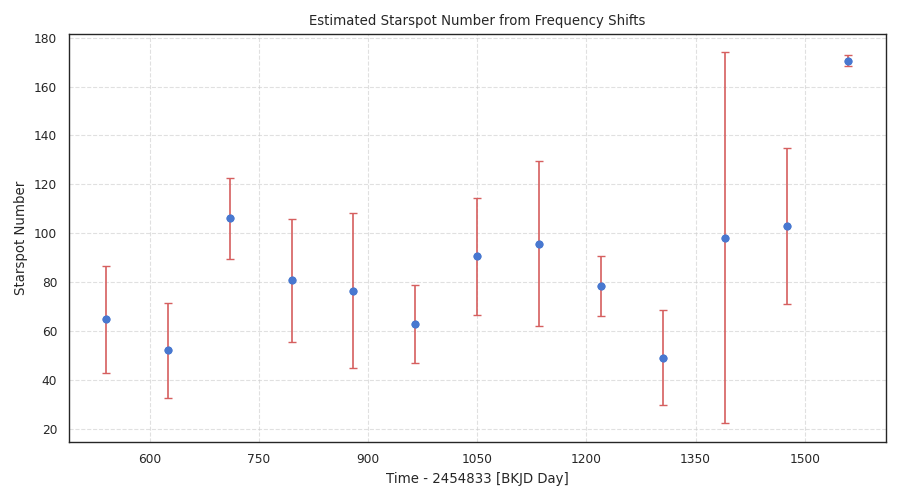}
\caption{Projected starspot number as a function of time.}
\label{starspots}
\end{figure}

\subsection{Comparison with KIC 6106415}
When comparing the asteroseismic oscillation frequencies of KIC 6106415 with those measured by GOLF, several notable similarities emerge. Higher-frequency modes in KIC 6106415 exhibit more pronounced temporal variations, a behavior also observed in the Sun. This suggests that similar physical processes—such as perturbations in the acoustic cavity due to surface magnetic fields and temperature fluctuations—govern frequency shifts in both KIC 6106415 and the Sun. However, the amplitude of these variations in KIC 6106415 is significantly larger than in the Sun. This difference likely reflects a more dynamic magnetic environment in KIC 6106415’s outer layers, which is consistent with the detection of frequent stellar flares in its light curve.

To further analyze these temporal variations, the oscillation frequencies of KIC 6106415 were divided into high-frequency and low-frequency regimes, using a threshold midway between the observed minimum and maximum frequencies. The high-frequency modes, which are more sensitive to the star’s outer convective layers, displayed larger temporal shifts, correlating with variations in magnetic activity. In contrast, the low-frequency modes, which probe deeper interior regions, exhibited smaller shifts, reinforcing the hypothesis that the primary drivers of these variations originate in the outer convective zones.

Additionally, the temporal evolution of KIC 6106415’s oscillation frequencies suggests the presence of a magnetic activity cycle. The periodic peaks and troughs in frequency amplitudes hint at a repeating pattern analogous to the solar cycle but operating on a shorter timescale, likely due to the star’s faster rotation rate. This supports the potential of using asteroseismology to infer magnetic activity cycles in solar-type stars.
\subsection{Estimating Starspot Numbers from Frequency Shifts}

$\mathit{Figure 13 }$ shows the how the starspot numbers vary according to the relative frequency shifts after averaging over all $\mathit{n}$ and $\ell$ values. The findings were consistent with expectation that the greater frequency variations aligned with greater sunspot numbers. After performing linear regression to extract the slope and intercept from this plot, a best fit line for estimating the starspot values was constructed according to
\begin{gather}
S = \delta\nu \cdot a + b
\end{gather}
where $\mathit{S}$ represents the estimated starspot number,$\mathit{\delta\nu}$ is the frequency shift, $mathit{a}$ is the slope derived from the regression fit, $\mathit{b}$ is the intercept.
Figure 14 shows the measured frequency shifts from the peakbagging results and the estimated starspot numbers from this equation. As expected, the largest shifts corresponded to a higher starspot number, consistent with the greater rotational period of the KIC 6106145.

\subsection{Conclusion}

Overall, the comparison between KIC 6106415 and the GOLF data underscores the universality of certain asteroseismic phenomena across solar-type stars, while also revealing intriguing differences in the magnitude and nature of magnetic activity. The larger frequency shifts observed in KIC 6106415 suggest that it may represent a different evolutionary stage or possess a more robust convective envelope, leading to stronger magnetic fields. These findings offer valuable insights into stellar dynamo processes and the complex interplay between magnetic activity and stellar oscillations in solar-type stars.

\begin{table}[h!]
\centering
\caption{Stellar Parameters for KIC 6106415}
\label{tab:kic6106415}
\begin{tabular}{lc}
\toprule
\textbf{Parameter} & \textbf{Value} \\
\midrule
Star Identifier & KIC 6106415 \\
Effective Temperature (Teff) & \SI{6037}{\kelvin} \\
Mass & \SI{1.039}{M_\odot} \\
Radius & \SI{1.213}{R_\odot} \\
Rotation Period & \SI{16.19}{\day} \\
Metallicity ([Fe/H]) & \SI{-0.04}{} \\
Numax & \SI{2249}{\micro\hertz} \\
DeltaNu & \SI{104.07}{\micro\hertz} \\
\bottomrule
\end{tabular}
\end{table}

\section{Acknowledgments}
We gratefully acknowledge the support of $\mathit{COFFIES}$ (Consequences of Fields and Flows in the Interior and Exterior of the Sun), a $\mathit{NASA}$ DRIVE Science Center (NASA Grant 80NSSC20K0602), for providing funding and resources that made this research possible. $\mathit{COFFIES'}$ dedication to advancing our understanding of stellar and solar dynamics has been instrumental in inspiring this investigation. We also extend our thanks to the $\mathit{GOLF}$ instrument team for their invaluable contributions to helioseismology through two decades of high-precision Doppler velocity data. Additionally, we acknowledge the $\mathit{Kepler}$ mission for its unparalleled dataset of stellar photometry, which has opened new frontiers in asteroseismology and stellar variability studies. These groundbreaking missions and collaborative efforts have collectively enriched our understanding of solar and stellar physics.

\facilities{NASA, ESA, COFFIES, GOLF, Kepler}

\bibliographystyle{aasjournal}

\begin{thebibliography}{}
\expandafter\ifx\csname natexlab\endcsname\relax\def\natexlab#1{#1}\fi
\providecommand{\url}[1]{\href{#1}{#1}}
\providecommand{\dodoi}[1]{doi:~\href{http://doi.org/#1}{\nolinkurl{#1}}}
\providecommand{\doeprint}[1]{\href{http://ascl.net/#1}{\nolinkurl{http://ascl.net/#1}}}
\providecommand{\doarXiv}[1]{\href{https://arxiv.org/abs/#1}{\nolinkurl{https://arxiv.org/abs/#1}}}

\bibitem[{Borucki {et~al.}(2010)Borucki, Koch, Basri, Batalha, Brown, Caldwell, Caldwell, Christensen-Dalsgaard, Cochran, DeVore, Dunham, Dupree, Gautier, Geary, Gilliland, Gould, Howell, Jenkins, Kondo, Latham, Marcy, Meibom, Kjeldsen, Lissauer, Monet, Morrison, Sasselov, Tarter, Boss, Brownlee, Owen, Buzasi, Charbonneau, Doyle, Fortney, Ford, Holman, Seager, Steffen, Welsh, Rowe, Anderson, Buchhave, Ciardi, Walkowicz, Sherry, Horch, Isaacson, Everett, Fischer, Torres, Johnson, Endl, MacQueen, Bryson, Dotson, Haas, Kolodziejczak, Van~Cleve, Chandrasekaran, Twicken, Quintana, Clarke, Allen, Li, Wu, Tenenbaum, Verner, Bruhweiler, Barnes, \& Prsa}]{Borucki2010}
Borucki, W.~J., Koch, D., Basri, G., {et~al.} 2010, Science, 327, 977, \dodoi{10.1126/science.1185402}

\bibitem[{Breton {et~al.}(2022)Breton, García, Ballot, Delsanti, \& Salabert}]{Breton2022}
Breton, S.~N., García, R.~A., Ballot, J., Delsanti, V., \& Salabert, D. 2022, Astronomy \&amp; Astrophysics, 663, A118, \dodoi{10.1051/0004-6361/202243330}

\bibitem[{Broomhall {et~al.}(2015)Broomhall, Pugh, \& Nakariakov}]{Broomhall2015}
Broomhall, A.-M., Pugh, C., \& Nakariakov, V. 2015, Advances in Space Research, 56, 2706, \dodoi{10.1016/j.asr.2015.04.018}

\bibitem[{Demarque \& Guenther(1999)}]{Demarque1999}
Demarque, P., \& Guenther, D.~B. 1999, Proceedings of the National Academy of Sciences, 96, 5356, \dodoi{10.1073/pnas.96.10.5356}

\bibitem[{{Gabriel} {et~al.}(1995){Gabriel}, {Grec}, {Charra}, {Robillot}, {Roca Cort{\'e}s}, {Turck-Chi{\`e}ze}, {Bocchia}, {Boumier}, {Cantin}, {Cesp{\'e}des}, {Cougrand}, {Cr{\'e}tolle}, {Dam{\'e}}, {Decaudin}, {Delache}, {Denis}, {Duc}, {Dzitko}, {Fossat}, {Fourmond}, {Garc{\'\i}a}, {Gough}, {Grivel}, {Herreros}, {Lagard{\`e}re}, {Moalic}, {Pall{\'e}}, {P{\'e}trou}, {Sanchez}, {Ulrich}, \& {van der Raay}}]{Gabriel1995}
{Gabriel}, A.~H., {Grec}, G., {Charra}, J., {et~al.} 1995, \solphys, 162, 61, \dodoi{10.1007/BF00733427}

\bibitem[{GOLF(2020)}]{GOLF}
GOLF. 2020, GOLF Global Oscillations at Low Frequencies,  European Space Agency, \dodoi{10.5270/esa-ls55aku}

\bibitem[{Ilin(2021)}]{Ilin2021}
Ilin, E. 2021, Journal of Open Source Software, 6, 2845, \dodoi{10.21105/joss.02845}

\bibitem[{Kosovichev(2011)}]{Kosovichev2011}
Kosovichev, A.~G. 2011, Advances in Global and Local Helioseismology: An Introductory Review (Springer Berlin Heidelberg), 3--84, \dodoi{10.1007/978-3-642-19928-8_1}

\bibitem[{{Lightkurve Collaboration} {et~al.}(2018){Lightkurve Collaboration}, {Cardoso}, {Hedges}, {Gully-Santiago}, {Saunders}, {Cody}, {Barclay}, {Hall}, {Sagear}, {Turtelboom}, {Zhang}, {Tzanidakis}, {Mighell}, {Coughlin}, {Bell}, {Berta-Thompson}, {Williams}, {Dotson}, \& {Barentsen}}]{2018ascl.soft12013L}
{Lightkurve Collaboration}, {Cardoso}, J.~V.~d.~M., {Hedges}, C., {et~al.} 2018, {Lightkurve: Kepler and TESS time series analysis in Python}, Astrophysics Source Code Library.
\newblock \doeprint{1812.013}

\bibitem[{Lopes \& Gough(2001)}]{Lopes2001}
Lopes, I.~P., \& Gough, D. 2001, Monthly Notices of the Royal Astronomical Society, 322, 473, \dodoi{10.1046/j.1365-8711.2001.03940.x}

\bibitem[{{Mikulski Archive for Space Telescopes}(2023)}]{MAST2023}
{Mikulski Archive for Space Telescopes}. 2023, Kepler Data Archive, \url{https://mast.stsci.edu}

\bibitem[{Reinhold {et~al.}(2013)Reinhold, Reiners, \& Basri}]{Reinhold2013}
Reinhold, T., Reiners, A., \& Basri, G. 2013, Astronomy \&amp; Astrophysics, 560, A4, \dodoi{10.1051/0004-6361/201321970}

\bibitem[{Salabert {et~al.}(2011)Salabert, Garcia, Palle, \& Jimenez}]{Salabert2011}
Salabert, D., Garcia, R.~A., Palle, P.~L., \& Jimenez, A. 2011, Journal of Physics: Conference Series, 271, 012030, \dodoi{10.1088/1742-6596/271/1/012030}

\bibitem[{Salabert {et~al.}(2015)Salabert, García, \& Turck-Chièze}]{Salabert2015}
Salabert, D., García, R.~A., \& Turck-Chièze, S. 2015, Astronomy \&amp; Astrophysics, 578, A137, \dodoi{10.1051/0004-6361/201425236}

\bibitem[{VanderPlas(2018)}]{VanderPlas2018}
VanderPlas, J.~T. 2018, The Astrophysical Journal Supplement Series, 236, 16, \dodoi{10.3847/1538-4365/aab766}

\end{thebibliography}

\end{document}